\documentstyle[11pt,pasp,twoside,psfig]{article}
\def\edcomment#1{\iffalse\marginpar{\raggedright\sl#1\/}\else\relax\fi}
\reversemarginpar

\begin{document}
\title{The Future of Fe-K Line Diagnostics for Probing Strong Gravity}
\author{Tahir Yaqoob}
\affil{Department of Physics and Astronomy, Johns Hopkins University,
3400 N. Charles St., Baltimore, MD21218, USA}

\begin{abstract}
We review what we have learnt with {\it ASCA} from studying the
Fe-K lines in AGN and describe a program to deconvolve
the narrow, non-disk components of the lines with {\it Chandra}.
This is necessary to derive the correct profiles of the
broad, relativistic lines obtained using data from {\it XMM}
and other high-throughput instruments. Since reverberation
techniques are now not looking promising, we present
{\it Constellation-X} simulations showing an
alternative way we might be able to measure black-hole mass and spin. 
\end{abstract}

\section{The Fe-K Lines in AGN }

Given the potential of the Fe-K line in AGN to yield constraints on 
black-hole mass and spin (e.g. Fabian et al. 2000), 
what have we actually learnt so far?
One thing which has been clear 
since soon after the launch of {\it ASCA} in 1993,
is that there
is a large variety of line profiles. In some
objects the line is very broad while in others it is not so broad.
Obviously it is the broadest lines that have the potential
for measuring black-hole parameters since the broadening is
believed to be due to extreme Doppler and gravitational energy
shifts close to the hole. Unfortunately only a handful of AGN
are bright enough to yield {\it ASCA} data with sufficient 
signal-to-noise to provide useful constraints on models of the
Fe-K line profile generated in an accretion disk rotating around
a black hole. For these, time-averaged spectra can produce fairly tight 
constraints on the disk inclination angle but this parameter
is degenerate with the rest energy of the line. {\it Unless
the line is not very broad, we cannot constrain the rest
energy of the line}. Thus, even in MCG$-$6$-$30$-$15, we cannot
yet tell from the Fe-K line profile what the dominant species of Fe ions
in the disk are. Since we have little information on what
the radial distribution of Fe-K line emission from the disk is,
the shape of the line between the extreme energies of the profile does not
yet provide useful constraints. 

We do not even know whether the
line emission is axisymmetric. If the fluorescent Fe-K line is
energized by local flares on the disk (as opposed to a central
source), it is likely not to be axisymmetric.
Moreover, we can only constrain the inner radius in a few
cases. If the Fe-K line is not very broad we can get a
lower limit on the radius of the line-emitting 
region (below this the disk may be completely
ionized or non-existent). If
the Fe-K line is extremely broad or redshifted, we may deduce
an upper limit on the smallest radius contributing to line emission. 
On two occasions ($\sim 40$ ks episodes in {\it ASCA} observations
of MCG$-$6$-$30$-$15) it has been possible to deduce that 
the inner radius is inside the marginally stable orbit
of a Schwarzschild black hole. Whether this provides evidence
for a Kerr hole has been debated.
Robust and unambiguous, model-independent measurements of black-hole
spin, one of the ultimate goals of these studies, has not yet been
achieved.

Nonaxisymmetric events are also a likely factor
in accouting for another major
result that has come out of Fe-K line studies. That is, rarely
does the line intensity (and shape, when it can be measured)
respond to changes in the continuum in a way which would be expected
from a simple unionized uniform disk responding to the
luminosity variations of a single, localized continuum source,
or one which uniformly covers the whole disk
(e.g. Reynolds 2000; Vaughan \& Edelson 2001). 
Ionization of the disk could also play
a role but whether this is the case or not, reverberation
techniques to measure the black-hole mass and spin will
not be possible until we understand the physics of the line
variability. When we do, reverberation mapping of the disk
using the Fe-K line may be impractical even with 
{\it Constellation-X}.

\section{ Deconvolving the Narrow, Non-relativistic Fe-K Line}

\begin{center}
\vspace{-5mm}
\begin{table}
\begin{center}
\caption{The Narrow Fe-K Line Component in Some AGN}
\begin{tabular}{lccc}
\tableline
Source & $E_{\rm Fe-K}$ (keV) & EW (eV) & FWHM (km/s) \\
\tableline
NGC 5548 & $6.402 \ (+0.027,-0.025)$ & $133\ (+62,-54)$ & $<8040$ \\
NGC 3783 & $6.347\ (+0.050,-0.021)$ & $86 \ (+35,-28)$ & $<6665$ \\
NGC 4151 & $6.386 \ (+0.014,-0.016)$ & $330 \ (+63,-67)$ & $<8169$ \\
Mkn 509 & $6.389 \ (+0.082,-0.030)$ & $61 \ (+35,-21)$ & $<6290$ \\
NGC 4051 & 6.4 (fixed) & $135 \ (+57,-53)$ & - \\
NGC 3327 & 6.4 (fixed) & $<85$ & - \\
3C 273 & 6.4 (fixed) & $<25$ & - \\
\tableline
\tableline
\end{tabular}
\end{center}
\end{table}
\end{center}

The {\it Chandra} {\it HETG}, combined with a high-throughput instrument 
such as {\it EPIC/XMM}, or  {\it PCA/RXTE}, currently offers the best possibility of
deconvolving any component of the Fe-K line which originates
far from the central black hole. Since the EW of such a narrow component
can be a significant fraction of the total Fe-K line emission, it
is important to measure, in order to correctly model the
relativistic component (see Yaqoob et al. 2001; Reeves et al., these
proceedings). Table 1 shows some preliminary {\it Chandra} measurements
for some AGN (errors and upper limits are for 90\% confidence, one parameter;
energies are in the rest-frame). 
We have initiated a program to simultaneously measure the
Fe-K line profiles of the brightest AGN with the strongest
Fe-K lines using {\it Chandra} and/or {\it RXTE} and/or {\it XMM},
results of which will be forthcoming. 

Figure 1 shows the {\it HEG} data directly against
(non-simultaneous)
{\it ASCA} data for four of these sources. 
Note that in some cases, such as 3C 273, a narrow component
is weak or absent. It is likely
that the narrow component does not vary on timescales of weeks
or perhaps much longer. However, we do not know, and should 
attempt to constrain the variability with {\it ASTRO-E2} and even
{\it XMM} in some cases. Even if the narrow component varies,
folding the {\it Chandra} results back into non-simultaneous
{\it ASCA} (or {\it XMM}) data is still useful because it is
unlikely that the energy and width of the narrow Fe-K line varies.

\begin{figure}
\centerline{\psfig{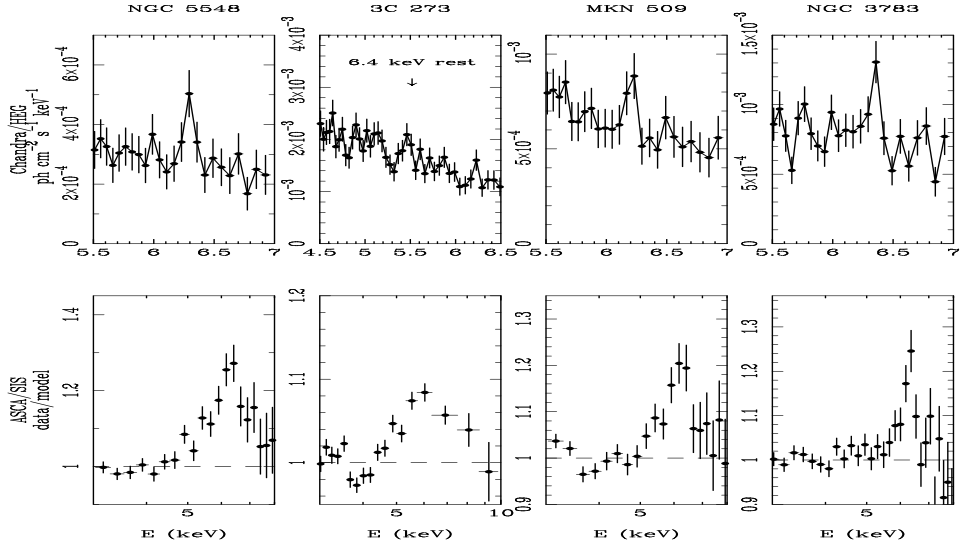}}
\caption{{\it Chandra  HETG} and {\it ASCA} Fe-K line profiles.
}\label{fig2}
\end{figure}

\section{ Measuring Black-Hole Spin}
Suppose there is an enhancement of the Fe-K line emission from the disk
at a few gravitational radii out from from the marginally stable orbit,
due to a local magnetic flare. Further suppose that this hot spot co-rotates
with the disk for at least one orbit. The result is that if we
examine the time-averaged line profile over the whole  disk, the
enhanced annulus will produce two very sharp spikes corresponding to
the extreme red and blueshifts. They are sharp because the annulus is
thin. The energies of these spikes are a function of the black-hole
spin ($a$), the inclination angle of the disk ($\theta$), and the
radius of the annulus ($r$). We can obtain  $\theta$ from the overall
line profile. That leaves two unknowns ($a$, $r$) and two observables
(the energies of the spikes). Hence we can measure $a$. The largest
uncertainty will be the rest energy of the Fe-K line.
Simulations show that {\it Constellation-X}
will easily be able to measure the energies of these spikes, superimposed
on the overall line profile. For example, Figure 2 shows a 40 ks simulation
of an AGN in which $r=6.5r_{g}$, $a=0$, and $\theta = 10^{\circ}$; the
2--10 keV flux is $9 \times 10^{-11} \rm \ erg \ cm^{-2} \ s^{-1}$. 
The EW of the overall Fe-K line is 300 eV and {\it the hot spot contains only
a mere 10\% of the total line emission}. How will we know whether we
have integrated over more than one orbit? We could split the observation, into
two, for example. The positions of the spikes should not change.
The Keplerian orbital timescale is
$314(r/r_{g})^{3/4} \ M_{7}$ s ($r_{g} \equiv GM/c^{2}$). 
For $M_{7}=1$, at $6r_{g}$ 
this is 4.6 ks and at $30r_{g}$ it is 50 ks. For $a>0$ these timescales
will be shorter, corresponding to smaller inner disk radii. 
\begin{figure}
\vspace{10pt}
\centerline{\psfig{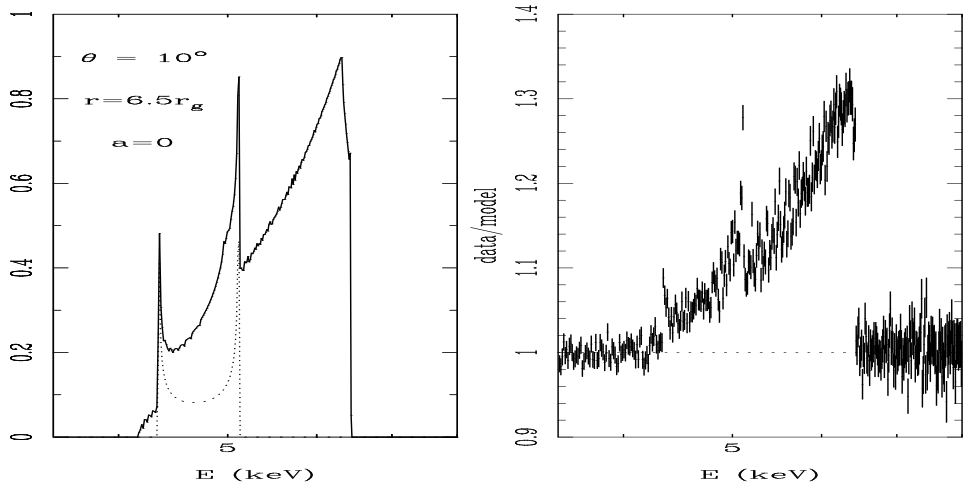}
}
\caption{40 ks {\it Con-X} simulation of a 10\% intensity hot spot. 
}\label{fig3}
\end{figure}
 
If such hot spots exist, the advantage of
this method for measuring spin is that it does not rely on having to
compute the detailed transfer function of the disk in
the face of complex ionization physics, nor on knowledge of
the line-emissivity function over the disk. 
Note that measuring the extrema of the main profile to obtain the
same information is not practical. Even future missions will have
trouble distinguishing the low-energy end of the Fe-K line from
the continuum. In contrast the narrow spikes will be easy to measure.
Time-resolved {\it ASCA} spectra
show evidence that these may exist, although the
bumps and wiggles which come and go are at the $2-3\sigma$ level
(e.g. Wang, Wang, \& Zhou 2001).
{\it XMM} should be able to pull them out of the noise if they are
real.

With {\it Constellation-X}, we could time-resolve the data into intervals
as small as the statistics allow and follow the energy of the narrow line
from the hot spot as it rotates, as suggested by Nayakshin \& Kazanas (2001). 
This will directly
constrain the black-hole mass. Sensitivity to this depends on the
size of the hot spot (i.e. width of the line). For example, taking
a Gaussian with $\sigma = 20$ eV, and the same parameters as in the
time-averaged model above, simulations show that $\sim 500$ s intervals with
{\it Constellation-X} will yield robust measurements of the line energy
in each time frame, with a 90\% confidence error of only 3 eV. 

I thank my collaborators, I. George, J. Turner,
K. Nandra, K. Weaver, \& J. Gelbord.

\end{document}